\documentclass[sigconf,screen]{acmart}

\AtBeginDocument{%
  \providecommand\BibTeX{{%
    \normalfont B\kern-0.5em{\scshape i\kern-0.25em b}\kern-0.8em\TeX}}}

\setcopyright{acmcopyright}
\copyrightyear{2018}
\acmYear{2018}
\acmDOI{10.1145/1122445.1122456}

\copyrightyear{2021} 
\acmYear{2021} 
\setcopyright{acmlicensed}\acmConference[CHI '21]{CHI Conference on Human Factors in Computing Systems}{May 8--13, 2021}{Yokohama, Japan}
\acmBooktitle{CHI Conference on Human Factors in Computing Systems (CHI '21), May 8--13, 2021, Yokohama, Japan}
\acmPrice{15.00}
\acmDOI{10.1145/3411764.3445720}
\acmISBN{978-1-4503-8096-6/21/05}

\usepackage{graphics}      
\usepackage[T1]{fontenc}   
\usepackage{txfonts}
\usepackage{mathptmx}
\usepackage{booktabs}
\usepackage{textcomp}
\usepackage{xcolor}

\usepackage{tabularx}
\usepackage{tabulary}
\usepackage{colortbl}
\usepackage{makecell}
\usepackage{multirow}
\usepackage{rotating}
\usepackage{enumitem}
\usepackage{color,soul,transparent}

\usepackage{microtype}        
\usepackage{ccicons}          
\usepackage{enumitem}

\begin{document}


\title{ The Public Life of Data:\\ Investigating Reactions to Visualizations on Reddit}


\author{Tobias Kauer}
\affiliation{
    \institution{The University of Edinburgh}
    \institution{University of Applied Sciences Potsdam}}
\email{t.kauer@ed.ac.uk}

\author{Arran Ridley}
\affiliation{
     \institution{University of Applied Sciences Potsdam}}
\email{arran.ridley@fh-potsdam.de}

\author{Marian Dörk}
\affiliation{
     \institution{University of Applied Sciences Potsdam}}
\email{marian.doerk@fh-potsdam.de}

\author{Benjamin Bach}
\affiliation{
    \institution{The University of Edinburgh}}
\email{bbach@ed.ac.uk}

\renewcommand{\shortauthors}{Kauer et al.}

\newcommand{\tobi}[1]{\textcolor{cyan}{#1}}
\newcommand{\tobicomment}[1]{\textcolor{cyan}{\textit{\textbf{Comment:} #1 [tk]}}}
\newcommand{\arrancomment}[1]{\textcolor{teal}{\textit{\textbf{Comment:} #1 [ar]}}}
\newcommand{\arran}[1]{\textcolor{teal}{#1}}
\newcommand{\marian}[1]{\textcolor{purple}{ \textit{#1 [marian]}}}
\newcommand{\bennew}[1]{\textcolor{red}{#1}}
\newcommand{\ben}[1]{\textcolor{blue}{\textit{#1 [ben]}}}

\definecolor{reaction}{HTML}{FFFFFF}
\definecolor{motivation}{HTML}{DAEEF2} 
\DeclareRobustCommand{\reaction}[1]{{\sethlcolor{reaction}\hl{\textit{``#1''}}}}
\DeclareRobustCommand{\motivation}[1]{{\sethlcolor{motivation}\hl{\textit{``#1''}}}}

\newcommand{\vistag}[1]{{\textsc{[#1]}}}

\begin{abstract}
This research investigates how people engage with data visualizations when commenting on the social platform Reddit. There has been considerable research on collaborative sensemaking with visualizations and the personal relation of people with data. Yet, little is known about how public audiences without specific expertise and shared incentives openly express their thoughts, feelings, and insights in response to data visualizations. Motivated by the extensive social exchange around visualizations in online communities, this research examines characteristics and motivations of people’s reactions to posts featuring visualizations. Following a Grounded Theory approach, we study 475 reactions from the \texttt{/r/dataisbeautiful} community, identify ten distinguishable reaction types, and consider their contribution to the discourse. A follow-up survey with 168 Reddit users clarified their intentions to react. Our results help understand the role of personal perspectives on data and inform future interfaces that integrate audience reactions into visualizations to foster a public discourse about data.

\end{abstract}

\keywords{data visualization, reddit, personal data}



\maketitle

\section{Introduction}
\label{chap:introduction}

The growing familiarity with data visualization and the emergence of easily accessible visualization authoring tools is fundamentally changing the ways in which visualizations are created, shared, and used. Today, visualizations are published through a wide range of channels, including news articles~\cite{segel2010narrative}, social media~\cite{shu2020makes}, discussion websites, blogs, reports, government websites,\footnote{\url{https://www.ons.gov.uk}} and  museums~\cite{hinrichs2008emdialog}. With this development, visualizations reach an ever widening and increasingly diverse range of audiences with their individual opinions, perspectives, and understandings. Learning how these audience engage with visualizations is key to understand how visualizations are used, how they can cause harm, and how they can contribute to productive discussions and informed decision-making. 

\textit{Social data-analysis}~\cite{wattenberg2005baby} describes the process of collaborative sensemaking with visualizations in online spaces. 
Many tools and platforms have been created for social data analysis, allowing for annotations, comments, and discussion threads~\cite[e.g.,][]{heer2007voyagers,viegas2007manyeyes,willett2011commentspace,wattenberg2005baby,walny2020pixelclipper}. Consequently, some investigations~\cite{heer2007voyagers,hullman2015content,wattenberg2005baby,elias2012annotating} have started reporting on peoples' comments and engagement with these platforms, reporting on how user comments refer to context, express criticism, and create a shared space for insights. While collaborative sensemaking is often associated with domain expertise and a commonly shared goal to explore data and generate insights~\cite[e.g.,][]{heer2007voyagers,walny2020pixelclipper,willett2011commentspace}, much engagement with visualization is happening without specialized or shared domain expertise or common goal.

Here, we are interested in how these `general audiences' engage with visualizations online. As our study suggests, engagement of these non-expert audiences can be very different and include motivations beyond (collaborative) sensemaking such as entertainment~\cite{sprague2012exploring}. There is much more to learn about how people engage with visualization in the broadest sense, including aspects such as critique~\cite{hullman2015content}, emotions~\cite{kennedy2018feeling}, personal experience~\cite{peck2019data} and expertise, and potentially visualization literacy. A better understanding of these social practices can inform the design of a future generation of annotation and collaboration tools as well as methods to allow larger and more diverse audiences to express their viewpoints and engage in collaborative exchanges and decision making with visualization. 

To widen our understanding of how people engage with data visualizations in the wild, our paper analyzes user comments on the Reddit forum \texttt{/r/dataisbeautiful}. We study community members' expressions and introduce the term \textit{reaction} to describe these expressions; we define a reaction as \textit{any verbal or textual expression upon viewing, reading, and/or interacting with a data visualization.} Through iterative, multi-person coding of 475 comments, based on principles of Grounded Theory, we obtained a classification of 10 distinct reactions types: \textit{observations, hypotheses, opinions, conclusions, clarifications, proposals, critiques, additional information, testimonies}, and \textit{jokes} (Section \ref{sec:reactions}). Reactions were found to point to four main \textit{scopes} around these visualizations: \textit{visual representation, data and data analysis, topic}, and \textit{insights}. 

These types and scopes provide us with a structured framework to analyze user's engagement described by our reactions. Our findings highlight a broad and diverse range of engagement types (reactions) and their respective subflavors, complementing prior research ~\cite{hullman2015content,wattenberg2005baby}. An online survey with 168 posters of comments from our collection delivers additional information about the audience's demographic and individual motivations (Section \ref{sec:personalpublic}). We found motivations to include both personal (intrinsic) goals and extrinsic goals to actively contribute to a public discourse around data. Finally, we discuss how our findings can inform the design of future generations of tools for collaborative and public discourses based on data visualizations (Section \ref{sec:discussion}).
\section{Background}
\label{chap:Background}

This research builds on prior studies of engagement with data visualization and collaborative sensemaking as well as recent research on personal data practices and user annotations.

\subsection{Engagement with Data Visualization}

The different aspects of the relationship between a visualization and its audience are often discussed in visualization as \textit{engagement}, a term that however lacks a firm definition. For example, Haroz et al.~\cite{haroz2015isotype} refer to engagement as a measurement of effectiveness when interacting with visualization, alongside memorability or performance. Other sources describe engagement as a prerequisite for learning, ranging from \textit{not-viewing} over \textit{responding} to \textit{presenting}~\cite{naps2002exploring}. Similarly, Mahyar et al. describe engagement by the means of Bloom's taxonomy of learning goals, ranging from \textit{low} to \textit{high}~\cite{mahyar2015towards}; \textit{low} has been attributed to viewing and simple interaction, while \textit{high} has been used to describe higher-level cognitive tasks such as hypothesis testing and decision making. Similar notions to describe engagement refer to \textit{``the [whole] processes of looking, reading, interpreting and thinking that take place when people cast their eyes on data visualizations and try to make sense of them.''}~\cite{kennedy2016engaging}.

From another perspective, engagement has been used to describe how visualizations can get citizens informed about and involved in social issues~\cite{dork2013critical}. In that sense, engagement is a process that may be triggered by a visualization, but eventually happens outside of the visualization, e.g., in the form of social change and advocacy, rather than sensemaking and interaction. Often such engagement will include other persons. Against the background of co-located and in-person conversations about visualizations, Walny et al. \cite{walny2020pixelclipper} provide a crucial differentiation between \textit{direct engagement}, which is \textit{``characterized by a visitor's observable interactions with the space, artifacts, and people''}, and \textit{data engagement}, defined \textit{``by the extent to which a visitor connects with [...] the data behind the visualizations''}~\cite{walny2020pixelclipper}. 

Developing a framework of different reaction types can
Some user engagement can be measured more easily, e.g,. from interaction logs, questionnaires, and observations~\cite{hung2017assessing,thudt2012bohemian,walny2020pixelclipper,conlen2019capture}. For example, prior studies \cite{thudt2012bohemian,wang2019comparing, valkanova2014myposition} report on observations on how general audiences engage with visualizations in co-located and real-world settings. However, rather than aiming to collecting a wide range of reactions, these studies focus on understanding positive and negative impressions towards a designed artefact such as an interactive table top application, a data comic, or a polling display. Data engagement and engagement beyond the artifact, on the other side, is hard to measure and assess. By relying on comments that users provide on the web, we explore a specific resource to understand engagement, while at the same time being objective and neutral about the type of engagement. Our reaction types are meant to 
broaden our understanding of engagement and can be seen as a specific form of \textit{visualization engagement} with data and visualization, highlighting forms of \textit{data engagement}.

\subsection{Collaborative Sensemaking}
To support engagement with visualization beyond viewing and exploration, a range of tools and interfaces have been built, supporting \textit{social data analysis}~\cite{wattenberg2005baby} and \textit{collaborative visualization}~\cite{isenberg2011collaborative}. These tools provide the audience with annotation capabilities ranging from free form annotation (e.g., text, pen-annotations) to specific pointers or recommending annotations~\cite{elias2012annotating}. For example,  CommentSpace~\cite{willett2011commentspace} introduced a fixed set of tags (e.g., \textit{question, hypothesis, to-do, evidence-for, evidence-against, etc.}) for authors to classify their own comments. 

Comments can also be linked to specific data points, providing evidence-for or evidence-against the given claims. Building on the idea of \textit{anchored conversations}~\cite{churchill2000anchored}, the sense.us platform by Heer et al.~\cite{heer2007voyagers} demonstrates a mechanism for linking comments with specific parts and states of a visualization, an idea recently extended by PixelClipper~\cite{walny2020pixelclipper}. The visualization community platform ManyEyes~\cite{viegas2007manyeyes} allowed users to upload data, create visualizations, share the visualizations online, and exchange comments with other community members. Finally, ChartAccent provides annotation functionality to support storytelling, available only to the authors but not to collaborators and audiences~\cite{ren2017chartaccent}. In general,  most of these tools, except ManyEyes, target specific groups of people, with specific tasks, and a common goal of sensemaking but little is known what people think beyond the features of these tools. To not introduce any functional bias into our research, we look only at textual comments while seeking to study an audience as large and diverse as possible.

\subsection{Studying User Annotations}

User annotations and comments have been studied in controlled user setups with both, business intelligence experts~\cite{elias2012annotating} and non-expert audiences~\cite{vanhulst2018descriptive, vanhulst2018designing}. An analysis of 300 annotations sourced though a lab study~\cite{vanhulst2018descriptive, vanhulst2018designing}, identified six dimensions of comments (\textit{insight on data, multiple observations, data units, level of interpretation, co-reference, detected patterns}). However, these studies investigated annotation features with a goal of collaborative sensemaking or evaluated specific commenting features of the designed tools. Moreover, as a controlled laboratory setup, these study results do not capture intrinsic motivations and reactions, such as personal traits, or expressions of opinions or feelings.

Wattenberg~\cite{wattenberg2005baby} reports on user comments made across a variety of social media while studying reactions to his BabyName Voyager. He points out the benefit of being able to point to evidence in a visualization when sharing one's comment and notes that a personal perspective can act as a simple entry point for people to engage with a visualization.

%
%
%
To the best of our knowledge there are only two systematic studies investigating reactions to visualization in online communities. Analyzing comments on a visual news website, Hullman et al.~\cite{hullman2015content} found that \textit{``comments drew on personal knowledge, opinions, experiences, or narratives''}. Analyzing comments on their Sense.us visualization commenting platform, Heer et al.\cite{heer2007voyagers} present a classification of annotations. However, annotations were created in controlled studies as well as intranets, while our work considers a vast public participation. Their classification eventually focuses on sensemaking categories such as \textit{observations, hypotheses, data integrity}, and social feedback (e.g., \textit{socializing}, \textit{affirmation}). In contrast, our work identified additional types of reactions such as critique and personal relations to data. Thus, we expand on these observations and devise a clear typology of reactions alongside a deeper analysis of personal drivers for engagement and public discourse.

\subsection {Personal Relations with Data}

A person's particular situation, knowledge, and experience is highly influential when engaging with a data visualization~\cite{kennedy2016engaging,kennedy2018feeling}. People's cultural and social context as well as the emotional condition play a considerable role when making sense of data. Any type of engagement can be related to current affairs and highly personal as reported by Peck et al.\cite{peck2019data}; for example, well-established visualization tasks (e.g., information extraction, hypothesis-building) seem to be overshadowed by biographical or personal factors; people would not trust data that is sourced from institutions with a political identity different from their own---no matter the topic or insight. In other cases, people would deem charts more useful if they depicted a topic they had first-hand experience with---independent of the extracted information. In this paper we investigate the notion that \textit{data is personal} and want to understand how personal relations to data are expressed in public settings. 
\section{Data \& Methodology}

For our study, we chose to investigate \href{https://reddit.com/}{Reddit} for its abundance and variety of data visualization posts, the lively discussions in response to them, and the ready availability of the data for qualitative analysis. Reddit has 430 million monthly active users (\textit{redditors})\footnote{\url{https://archive.is/w04Nx}}---for comparison, Twitter has 336 million\footnote{\url{https://www.statista.com/statistics/282087/number-of-monthly-active-twitter-users/}}---who share media, links or self-created content in groups (\textit{subreddits}) that focus on a broad range of cultural, political, or social topics. It has been conceptualized as a public sphere in which people engage collaboratively to express and form opinions~\cite{dosono2017exploring}. Contrary to other social platforms like Twitter or Facebook, Reddit users remain semi-anonymous, do not form networks~\cite{kilgo2016led}, and do not have ``an a priori built social network''~\cite{medvedev2017anatomy}. Posts are generally public and can be commented on in hierarchical threads by anybody. Posts and reactions to them receive a score by the audience (\textit{karma}), making reddit ``the internet's largest social voting community''~\cite{gilbert2013widespread}. 

Our study investigates visualization reactions expressed on the visualization-focused subreddit \texttt{\href{https://www.reddit.com/r/dataisbeautiful/}{/r/dataisbeautiful}}. What kind of content is allowed in this community is specified by a set of rules, stating that all posts must contain a data visualization that has at least one computer generated element. The rules also enforce a standard by sanctioning sensationalized headlines (\textit{``Clickbait posts will be removed''}) and encouraging the transparent display of the data sources. An additional set of rules is aimed at the audience to establish a code-of-conduct, inviting constructive feedback to the presented visualization and prohibiting intentionally rude or hateful comments. 
To extract data and content from Reddit, we use the official API through the Python library PRAW.\footnote{\url{https://praw.readthedocs.io/en/latest/index.html}}

\begin{table}
{\footnotesize
\begin{tabularx}{\linewidth}{p{0.2\linewidth}|X}
\texttt{blockbuster} & \href{https://reddit.com/r/dataisbeautiful/comments/hdrnfw/oc_blockbuster_video_us_store_locations_between/}{Blockbuster Video US store locations between 1986 and 2019} \\ \hline
\texttt{breaking-bad} & \href{https://reddit.com/r/dataisbeautiful/comments/fwjces/oc_the_absolute_quality_of_breaking_bad/}{The absolute quality of Breaking Bad.} \\ \hline
\texttt{bushfire} & \href{https://reddit.com/r/dataisbeautiful/comments/ejhw7w/area_of_land_burnt_in_australia_and_area_of_smoke/}{Area of land burnt in Australia and area of smoke coverage shown as equivalent area over Europe } \\ \hline
\texttt{cases} & \href{https://reddit.com/r/dataisbeautiful/comments/fj7535/oc_covid19_spread_from_january_23_through_march/}{COVID-19 spread from January 23 through March 14th. (Multiple people independently told me to post this here)} \\ \hline
\texttt{daylight} & \href{https://reddit.com/r/dataisbeautiful/comments/duax05/oc_hours_of_daylight_as_a_function_of_day_of_the/}{Hours of daylight as a function of day of the year and latitude} \\ \hline
\texttt{deaths} & \href{https://reddit.com/r/dataisbeautiful/comments/fxoxti/coronavirus_deaths_vs_other_epidemics_from_day_of/}{Coronavirus Deaths vs Other Epidemics From Day of First Death (Since 2000) } \\ \hline
\texttt{donations} & \href{https://reddit.com/r/dataisbeautiful/comments/hdaf6n/oc_top_10_highest_covid19_donations_with_the/}{Top 10 Highest Covid-19 donations with the percentage of their net worth} \\ \hline
\texttt{employment} & \href{https://reddit.com/r/dataisbeautiful/comments/fpga3f/oc_to_show_just_how_insane_this_weeks/}{To show just how insane this week's unemployment numbers are, I animated initial unemployment insurance claims from 1967 until now. These numbers are just astonishing.} \\ \hline
\texttt{garbage} & \href{https://reddit.com/r/dataisbeautiful/comments/cvoyti/the_great_pacific_garbage_patch_oc/}{The Great Pacific Garbage Patch } \\ \hline
\texttt{golf} & \href{https://reddit.com/r/dataisbeautiful/comments/ho1sh3/hardest_golfing_president_oc/}{Hardest Golfing President } \\ \hline
\texttt{google} & \href{https://reddit.com/r/dataisbeautiful/comments/hs9mnz/oc_trending_google_searches_by_state_between_2018/}{Trending Google Searches by State Between 2018 and 2020} \\ \hline
\texttt{IMDb} & \href{https://reddit.com/r/dataisbeautiful/comments/fqqzki/worst_episode_ever_the_most_commonly_rated_shows/}{Worst Episode Ever? The Most Commonly Rated Shows on IMDb and Their Lowest Rated Episodes } \\ \hline
\texttt{minecraft} & \href{https://reddit.com/r/dataisbeautiful/comments/efvgve/where_is_each_ore_found_in_a_minecraft_world_oc/}{Where is each ore found in a minecraft world? } \\ \hline
\texttt{net-worth} & \href{https://reddit.com/r/dataisbeautiful/comments/f6stjs/oc_2020_presidential_candidates_by_net_worth/}{2020 Presidential Candidates by Net Worth} \\ \hline
\texttt{oil} & \href{https://reddit.com/r/dataisbeautiful/comments/g5b14x/oc_us_oil_price/}{US oil price} \\ \hline
\texttt{ramsey} & \href{https://reddit.com/r/dataisbeautiful/comments/hr1131/oc_what_percent_gordon_ramsays_kitchen_nightmares/}{What percent Gordon Ramsay’s Kitchen Nightmares are still open?} \\ \hline
\texttt{rates} & \href{https://reddit.com/r/dataisbeautiful/comments/fhykic/oc_this_chart_comparing_infection_rates_between/}{This chart comparing infection rates between Italy and the US} \\ \hline
\texttt{stimulus} & \href{https://reddit.com/r/dataisbeautiful/comments/fppc7v/oc_where_the_money_goes_in_the_us_senates_2t/}{Where the money goes in the US Senate's \$2T coronavirus stimulus bill} \\ \hline
\texttt{tuition} & \href{https://reddit.com/r/dataisbeautiful/comments/hni7zy/us_college_tuition_fees_vs_overall_inflation_oc/}{US College Tuition \& Fees vs. Overall Inflation } \\ \hline
\texttt{tulsa} & \href{https://reddit.com/r/dataisbeautiful/comments/hfk5f4/oc_attendance_at_donald_trumps_rally_in_tulsa/}{Attendance at Donald Trump’s rally in Tulsa, compared to the number of tickets Trump claimed were requested.}
\end{tabularx}
\caption{Original titles of the visualization posts from which the reactions were obtained. We use short tags to identify the visualizations throughout this paper.}
 \label{Tab:vis-overview}
}
\vspace{-3em}
\end{table}

\subsection{Data Selection}

To find a suitable subset of comments for this study, we followed the following five-step procedure:

\begin{enumerate}[noitemsep,leftmargin=*]
    \item \textbf{Pilot studies:} We examined single visualization posts and their reactions and tested preliminary coding schemes on small numbers of them. In these pilots, we observed that some posts, e.g. visualization of quantified-self data and humorous posts, attract many inappropriate or off-topic reactions.
    \item \textbf{Visualization selection:} We included the 26 most popular visualization posts according to their \textit{karma} score over the span of one year (Oct-2019 - Sept-2020) in our study. From these 26 posts, we manually removed six posts that mostly attracted off-topic reactions and did not consider them for data collection.
    \item \textbf{Data Collection:} We used the reddit API to collect 68.000 reactions from the remaining 20 visualization posts.
    \item \textbf{Filtering:} We filtered these reactions based on the following four criteria and ended up with a set of 792 reactions.
    \begin{itemize}[noitemsep]
        \item As we intended to look only at direct reactions to the visualizations, we only included first-level comments but no replies to any comments. 
        \item We ignored comments by the visualization author and by automated bots, identifiable through their username.
        \item We ignored comments that were deleted by the reaction authors or removed by moderators.
        \item We set a threshold of five upvotes for each comment to be considered, to remove comments off topic, illegible, or otherwise irrelevant.
    \end{itemize}
    \item \textbf{Sample Sizing:} Since the number of reactions differed along the 20 posts, we randomly selected 25 reactions per visualization. Since some visualizations had less than 25 reactions, our final dataset contains 475 reactions.
\end{enumerate}


Generally speaking, the visualizations we considered use common and well-known techniques such as barcharts, linecharts, or heatmaps. Nine visualizations are animated, the other eleven are static images. While the range of depicted topics is broad, we found several visualizations about CoViD-19, the 2020 presidential election or pop-cultural data like \href{https://www.imdb.com/}{IMDb}-ratings. \autoref{Tab:vis-overview} shows the list of visualizations included in our study, with titles and clickable links.

\subsection{Open Coding}
To obtain a high-level understanding of reactions, we engaged in open and close coding following Grounded Theory~\cite{strauss1994grounded}. Our goal was to find independent and non-overlapping concepts in the data, able to describe the range of users' reactions.

Our coding scheme developed over eight major iterations. Each iteration involved between two to four of the authors and refined both the set of comments and visualizations as well as our coding scheme. We started our open coding by independently coding a random set of comments. We then build a first scheme with four dimensions: \textit{i)} types of reactions, e.g., observation, critique, clarification, etc,
\textit{ii)} a user's stance (negative, neutral, positive),
\textit{iii)} their conviction (weak, neutral, strong), and 
\textit{iv)} any presented evidence (e.g., external, anecdotal). 

Throughout the early iterations, we observed that many of these dimensions depended very much on the subjective interpretations of the coders. Through several iterations of close coding, we removed dimensions and categories that led to misleading coding. Each new iteration was tested with new set of randomly selected comments to prevent us from fitting our scheme to a specific set of comments. In the final coding scheme, 20\% of the data was coded by all four authors and discussed to reach consensus. The four authors independently used this scheme to categorize comments and test, whether the assumed categories could exhaustively and without overlap characterize the actual data. We merged categories that were hard to discern from our data (e.g., observations and interpretations) and added new categories where comments did not fit any of the existing categories or to clarify important differences among comments within a category. 

Throughout coding, we observed that many comments included notions about the commenter's personal perspectives. While we started out with an individual reaction type `personal reference' many of these comments overlapped with other reaction types. We thus introduced an optional \textit{personal} flag to tease out the role of personal qualifications in the reactions. Many of these personal comments included subtle expressions as first-person pronouns (I, me, my) or indications of experience (\textit{as someone who [...]}), to articulate personal positions (\textit{in my opinion [...]}). 

\subsection{Final Scheme}
After our 8 major iterations, we converged towards a coding scheme with two dimensions: Ten \textit{reaction types} (\autoref{tab:reactiontypes}) and four \textit{reaction scopes} (\autoref{fig:type-vs-scope}). While we started out with the intention to only allow one type per reaction, we accepted that comments are often more nuanced and ambiguous to be classified as one type only. We attribute this to the very nature of human thought and expression. While we tried to minimize overlap between types of reactions as much as possible, we accepted some comments to contain two reaction types, rather than eliminating overlap.

\subsection{Online Survey}
We conducted a survey and contacted the authors of all 475 comments, 168 participants (35\%) of which completed it. In the survey questionnaire, we asked for basic demographic information and compared them to the overall average of the Reddit user base, reported on.\footnote{\url{https://www.pewresearch.org/internet/2013/07/03/6-of-online-adults-are-reddit-users}} 10\% of the participants in our survey were not cis-male (overall on Reddit:~33\%). Our two most prominent age groups were 19-29 year old (survey:~45\%, overall: 50\%) and 30-39 year olds (survey: 45\%;~overall:~9\%) The participants lived most of their lives in North America (58\%) or Europe (28\%). 39\% had a Bachelor's degree, 25\% had a Master's degree, 24\% had a high school diploma. 45\% stated that they had some formal training with data, statistics or visualization. We asked the participants about their motivation for their reaction and report back on the results in Sections 4 and 5.
\section{Visualization Reactions}
\label{sec:reactions}

\begin{figure*}[t]
  \centering
      \includegraphics[width=2\columnwidth]{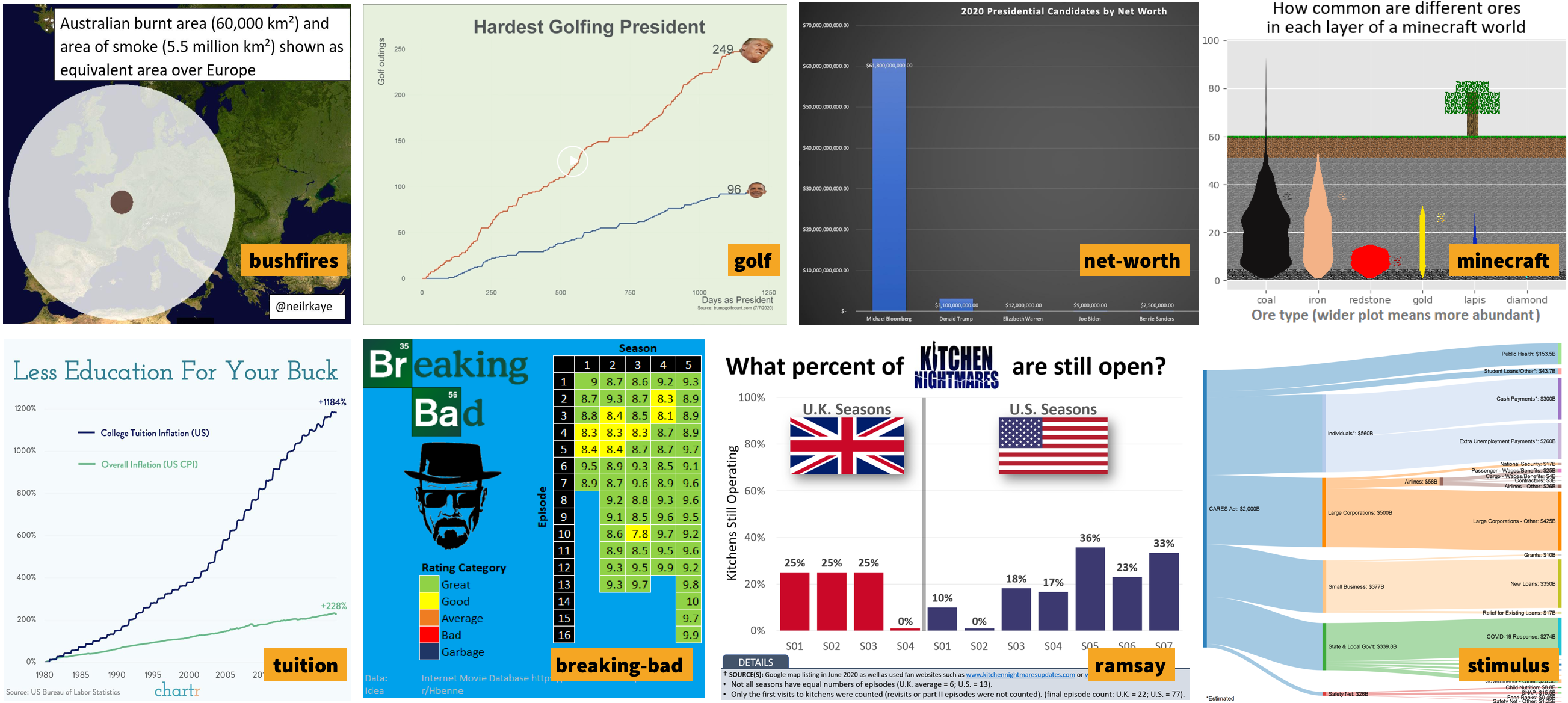}
  \caption{Examples of the 20 popular visualizations we obtained from \texttt{/r/dataisbeautiful}. In total we considered 475 reactions. Visualizations varied in their topics, aesthetic appeal, visualization techniques, and the use of animations.}~\label{fig:overview}
\vspace{-1em}
\end{figure*}

\begin{figure}[t]
\centering
\includegraphics[width=1\linewidth]{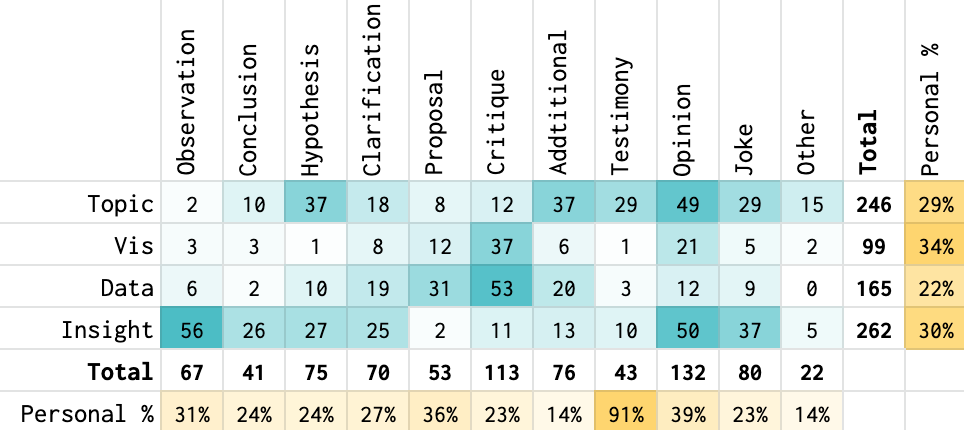}
\caption{A heatmap indicating the absolute distribution of types across scopes. Values that stick out are e.g., how critical reactions most often relate to the data, whereas the scope of observations most often is an insight. Yellow cells indicate the relative distribution of personal flags. By design, testimonies and opinions most often have a personal notion.}
\label{fig:type-vs-scope}
\end{figure}

This section presents ten reaction types and four scopes. The reaction types describe the kind of contributions that people make when responding to a visualization. The scopes describe which aspect of a visualization the comment refers to. In the following, we start by explaining scopes and then present the reaction types. We include verbatim examples of reactions and  show what visualization they refer to in a [\texttt{tag}]. We complement the results of our analysis with the motivations the posters shared in the survey. Direct quotes from the survey are indicated through a {\sethlcolor{motivation}\hl{\textit{background color}}.

\subsection{Scopes}
Akin to Hullman et al.'s~\cite{hullman2015content} targets, scopes denote the part of a visualization post a commenter refers to. Since posts on \texttt{/r/dataisbe\-au\-tiful} are structurally different from comments on a news blog and do not come with an accompanying article, we encountered four slightly different scopes:

\textbf{Data}---This scope describes reactions about the dataset. Comments referred to data quality and the ability to inform (\reaction{What is the composition of the large corporation other? Loans? Grants? Tax credits?} \vistag{stimulus}), data collection methods (\reaction{do you mind discussing how you parsed the world file?} \vistag{minecraft}), or the accuracy of data (\reaction{The swine flu killed at least 18000 people in the end. I find this quite misleading} \vistag{deaths}).  
    
\textbf{Visualization}---Comments with this scope refer to the visual representation and its creation. Comments referred to representation aesthetics (\reaction{Great Graphic} \vistag{stimulus}), visual variables (\reaction{Any meaning to the different colours?} \vistag{cases}), or other design decisions (\reaction{This map is using the Mercator projection, which can be a bit deceptive} \vistag{bushfire}). 

\textbf{Insight}---Whenever a comment referred to information that can be extracted from the visualization, we call this an `insight'. Comments referred to observed data points or trends (\reaction{It really revved up in the last 4 days.} \vistag{cases}, raised questions about a finding (\reaction{We can see where it goes.... but... where does it come from?} \vistag{stimulus}), or any other contemplation about the presented information (\reaction{Why is there a seasonal trend to this? What month does that happen and any idea why?} \vistag{employment}). 

\textbf{Topic}---Comments with this scope  referred to the topic illustrated by the visualization, rather than any specific insight. Generally, these comments were found to be independent from the specific information contained in a visualization: \reaction{We had so much time to start quarantining people travelling from China. We all blew it.} \vistag{cases}.

The majority of the 475 comments referred to either an insight (188) or to the topic (187) (\autoref{fig:type-vs-scope}). This supports Kennedy et al.'s~\cite{kennedy2016engaging} findings that a visualization's subject matter is an important driver for engagement---people engage with topics they are interested in. 117 reactions mainly referred to the data source, and 82 talked about the visual representation of that data.

\subsection{Reaction Types}
We now report on the ten reaction types that we identified. 
Each type indicates a distinct contribution that a comment makes in response to a data visualization. 
We provide a brief description, the frequency, and example reactions for each. We also discuss potential subgroups that are characterized by similar phrases or lines of argumentation. \autoref{tab:reactiontypes} provides an overview of the reaction types.

\textbf{Observations} point out insights that were made directly in the visualization or interpretations derived from its data: \reaction{wow, House episodes rating are incredible consistent} \vistag{IMDb}. These reactions consist of individual, often simple observations of outliers (\reaction{May 24th, when Minnesota became the center of the world for a little bit.} \vistag{google}) and trends (\reaction{Wow that's crazy I didn't think a company could expand that quickly. And then to see it disappear as quickly is crazy} \vistag{blockbuster}). Often, observations are a necessary prerequisite for other reactions, such as opinions and conclusions. Their scopes are predominantly insights, yet they are not merely about expressing what a person took away from a chart. Observations are often expressed in a humorous way and combine specific data points or trends with a joke: \reaction{That line dips any lower it'll hit oil} \vistag{oil}. We also found a number of observational reactions that include subjective or personal expressions, such as \textit{``I like'', ``I'm surprised'', ``I'm particularly amused by'', ``Fascinating that''}, and \textit{``Wow, that is [...]''}}. 
These expressions reveal genuine moments of surprise associated with finding something interesting and relevant. Moreover, they show people's personal stance with respect to that observation. We flagged these observations as \textit{personal} for subsequent analysis. 

\begin{table}[t]
\footnotesize
\begin{tabularx}{\linewidth}{p{0.19\linewidth}|X|p{0.01\linewidth}}
\textbf{Type} & \textbf{Description} & \textbf{\#} \\ \hline
\textsc{Observation} & Stating the obvious; point out specific data points or trends; express insights and interpretations& 62 \\ \hline
\textsc{Conclusion} & Taking a step back to provide an overview over the presented information and draw conclusions or call to action & 29 \\ \hline
\textsc{Hypothesis} & Providing a speculative explanation to explain the presented information. This can range from tentatively asked questions to bold claims that introduce additional data to argue for an explanation & 56 \\ \hline
\textsc{Clarification} & Asking for information to better understand the data, a visualization or its insight  & 50 \\ 
\hline
\textsc{Proposal} & Propose future work and possible adaptations of the data or visualization  & 43 \\ \hline
\textsc{Critique} & Feedback to the visualization author, ranging from expressing constructive feedback and pointing out improvements to disagreement with the visualization and contestation of data. & 90 \\ \hline
\textsc{Additional information} & Introduce additional information to allow people to better contextualize or relate to the presented data. Includes background information, comparisons with similar data, trivia, and  links to external sources with additional data. & 58 \\ \hline
\textsc{Testimony} & Providing personal information by speaking from first-hand experiences and sharing anecdotes or memories  & 41 \\ \hline
\textsc{Opinion} & Providing a personal perspective by sharing opinions, but also feelings and emotional reactions. & 92 \\ \hline
\textsc{Other} & Unfitting in other categories, off-topic or not understandable reactions & 19
\end{tabularx}
\caption{Reaction types, their meaning, and frequency of occurrence (\#). The sum of occurrence is larger than number of reactions due to some double encodings}
\label{tab:reactiontypes}
\vspace{-2em}
\end{table}

\textbf{Conclusions} take a step back and reflect on the observed information and its consequences.
One characteristic group of conclusive reactions simply summarize the presented information and refer to insights the reaction author had: \reaction{Season 5 was flawless, particularly those last 3 episodes [...]} \vistag{breaking-bad}. Beyond that, we found a number of comments that put individual observations into perspective: \reaction{When billions of dollars can be depicted using a thin, one pixel line, it truly shows just how much two trillion dollars is}~\vistag{stimulus}. Other reactions draw conclusions from the data to point out necessary next steps or make a call to action: \reaction{This is exactly why any push for free college MUST tackle the irresponsible spending and ballooning administrative costs by colleges} \vistag{tuition}. 
Conclusions can include a personal stance and can be motivated by expressing how this stance is not in line with the take-away message of the visualization: \motivation{I wanted to summarise the data in a different way to how the 
[author] did
}.

\textbf{Hypotheses} provide speculative explanations for the presented information. While some hypotheses are phrased as short, tentative questions (\reaction{Is it possible the only driver for cost of college is not inflation but rather demand?} \vistag{tuition}, the majority of them introduces additional information and argues for a specific explanation of the presented data: \reaction{[...] 
This perfectly tracks with the drive to transfer our publicly shared national treasure into the private accounts of the exceptionally wealthy, starting with Reagan in the 80's. This is wealth inequality come home to roost.
} \vistag{tuition}. Some of these claims introduce bold theories that are generally not externally verified: \reaction{This is all incentivized by the government [...]} \vistag{tuition}.

\textbf{Clarifications} are questions addressed to the visualization author or the public audience. They are generally posed to better understand the presented information. We found a wide variety of clarification questions: Some people needed help with correctly interpreting the information (\reaction{How is it negative, I don't know how these things work}, oil) or encoding the visualization (\reaction{What does the wideness of the color mean? [...]}
\vistag{minecraft}).
Some intend to get more details about specific data points (\reaction{What the hell happened in season 4 of the UK version?} \vistag{ramsey}) or background information about the topic (\reaction{Why is there a seasonal trend to this? What month does that happen and any idea why?} \vistag{bushfire}). One recurring clarification seeks more information about the used datasets and how they are generated: \reaction{Why is `nation security' part of the `large corperations branch?} \vistag{stimulus}. These clarifications are generally phrased in a more critical tone, as they question the design decisions that were made to produce the used data. This critical stance is also observable in some skeptical questions that question the correctness of single data points: \reaction{[...] your chart is showing that both poles go from 0 hours of daylight to 24 hours of daylight in what appears to be a single day. See day 82 for example. You sure your math is right?} \vistag{daylight}. Such questions often co-occur with the reaction type critique and are a frequent case of double-encoding.

\textbf{Proposals} suggest future work and possible amendments to the dataset and adaptations of the visualizations. While we observed different kinds of proposals in the reactions, a majority of proposals suggested to introduce additional data to the existing visualization to provide more context: \reaction{Can we add the common flu to this and start on January 1st?} \vistag{deaths}. Others were interested in using the same visualization to display another related dataset: \reaction{Can I get one of these for every spending bill? [...]} \vistag{stimulus}. On the other hand, reactions proposed to improve or change the visualization: \reaction{Kinda wish the opening and closing were more distinctively different} \vistag{blockbuster}, sometimes with the intention to make the presentation more relatable: \reaction{Please do British Columbia Canada! My wife doesn't get how bad and big it is} \vistag{bushfire}. One participant described the motivation to react with a proposal with the \motivation{Desire for more than one perspective on a dataset}.

\textbf{Critique} gathers reactions that express a form of resistance. In contrast to proposals, they identify specific issues with the current form of visualization or data---often in a harsh or ironic speech. The most predominant kinds of critique are contestations that scrutinize the validity of the used data for the visualization: The scope of 52 (out of 113, 46\% XX) critical reactions is on the data. These data contestations are threefold: Firstly, people point out missing data that they deem relevant to get a whole picture of the presented information: \reaction{Should have included cancer and AIDS to humble down those COVID stats} \vistag{deaths}. Secondly, people indicate that the presented data does not accurately represent the topic and should be generated in another way: \reaction{It seems that this graphic uses worldwide statistics for Covid19, but USA only stats for some of the other outbreaks, in order to give an overly dramatic impression} \vistag{deaths}. Thirdly, people contest the correctness of single data points, sometimes 
by introducing additional sources: \reaction{This is wrong. Based on the Deutsche Bank info that came out last week. Trump is only worth about 600-700 M. [...]} \vistag{net-worth}. This last point illustrates how contestations often co-occur with the provision of additional information. 
The motivation behind this kind of critique often is \motivation{to combat misleading or unhelpful data visualizations}.
Beyond these contestations of data we observe 36 (32\% XX)  criticisms of the visualization. The reactions point out hard to understand or possibly misleading representations (\reaction{This chart succeeds at illustrating the wealth of Bloomberg, but does a very poor job of displaying/contrasting most of the candidates. Consider a log plot} \vistag{net-worth} and often co-occur with suggestions for improvement.
Surprisingly many reactions address the aesthetic appeal of visualizations and we often observed harsh criticism for a perceived lack of beauty or the use of animation in charts.

\textbf{Additional information} is provided in order to contextualize the visualization. This can be background knowledge that helps to understand the subject (\reaction{[...] 
The garbage patch is not a big island of floating plastic bottles, etc. It's a big sparse conglomeration of micro plastics
[...]} \vistag{garbage} or to put the presented data into perspective (\reaction{Instead we had 13 million people watch the rally} \vistag{tulsa}). Since additional information is used to explain existing or introduce new knowledge, we observe that these reactions are among the longest in our study (>500 characters on average), with them being double as long as the overall average.
People's motivation to provide additional information is to contribute to sensemaking by adding what they deem relevant: \motivation{I had information pertaining to the topic at hand and just wanted to point out that the data was more significant.}
Others also have the discussion that can unfold from this kind of reaction in mind: \motivation{[...] 
i decided to post it to add the idea to the conversation}.


\textbf{Testimonies} are reports of first-hand experiences with the presented information, usually with a strong focus on the topic (70\%).
Testimonies express forms of local knowledge and anecdotal evidences of people's everyday lives, oftentimes disclosing biographical or private information. The largest group of testimonies talk about a personal spatial relation to the presented information: \reaction{I live along the east coast of Australia in an area not directly affected by the fire but heavily impacted by the smoke [...]} \vistag{bushfire}; \reaction{We don't have daylight savings here in India. Now I know why that's a phenomenon in the West. [...]} \vistag{daylight}. Others relate to the topic through memories and experiences they have made in the past: \reaction{I remember being a kid wanting to work at block buster one day. They were gone by the time I was old enough to work} \vistag{blockbuster}. A third group talks about the impact the presented data has on their everyday lives and disclose information about their jobs or families: \reaction{[...] I'm a little pissed I'm an essential worker throughout this and will be getting less pay than being unemployed [...]} \vistag{stimulus}. The last example shows that testimonies frequently co-occur with opinions. 

\textbf{Opinions} provide a personal perspective by sharing an individual judgement or point-of-view. This can be simple feedback to the visualization author (\reaction{I really enjoy this graph. It really bothers me when people conflate millionaires and billionaires} \vistag{net-worth}; \reaction{This data is most certainly not beautiful} \vistag{rates}). Often, opinions carry a notion of personal account, yet they do not disclose the same level of personal information as testimonies: \reaction{I don't think I will ever get or understand the appeal to play golf. Played once and was bored out of my mind} \vistag{golf}. Visualizations that depict everyday topics many people can connect with---such as pop-cultural data about TV shows---show a high number of opinionated reactions. On the one hand, people have opinions about the topic, people who watched the depicted TV show express how they liked it (\reaction{Glad to see Person of Interest get some love. Always felt that was an underrated show.	} \vistag{IMDb}. However, we also found some contestations when people relate their own experience to the data and express an opinionated contestation: \reaction{How they gunna hate on Towlie? That's one of my favorite episodes!} \vistag{IMDb}. This is backed up by the motivation to create these kinds of posts: \motivation{I liked the show. I didn't agree with the visualization}. This tension between an insight on the data and the commenter's own stance on that matter is a recurring theme we found in many opinionated reactions: \reaction{Might as well call this 'Why I hate people: An animated explanation' } \vistag{google}. 


How many redditors does it take to change a light bulb? Nine to express all prior types and one to make \textbf{jokes} about it. Across all visualizations, we observe many reactions that are primarily intended to be funny, including sarcasm (\reaction{ah yes, the two presidents of the united states} \vistag{golf}), pop-cultural references (\reaction{Game of Thrones' ending is why I have trust issues.} \vistag{cases}), observational humour (\reaction{Looks like every Facebook party ever organized} \vistag{tulsa}), and puns (\reaction{This chart points out the ultimate irony... It could not break into Bad...} \vistag{breaking-bad}). 
19 reactions we collected would not fit into any of the presented categories. This includes verbal insults, automated responses from bots or things we could not read or understand, e.g., \reaction{Meanwhile a mysterious diplomat traverses his homeland searching unsuccessfully for a single hand to shake} \vistag{bushfires}.

\section{The Personal and the Public: Why People Engage with Visualizations Online}
\label{sec:personalpublic}

While reaction types inform us about \textit{what} type of engagement people shared, this section looks into the \textit{why}, i.e., asking for the motivations behind people's comments. Based on the 168 survey replies we obtained from from comments, we report on findings about personal (intrinsic) motivations and motivations to support the public discourse (extrinsic).

\subsection{Personal Perspectives and Experiences}
A major theme that we observed when coding the reactions was that personal perspectives were expressed in many comments and throughout all types of reactions: testimonies, opinions, emotions (e.g., in observations), adding personal knowledge (e.g., in additional information), proposals, and critiques. Building on prior work on personal perspectives in data visualization~\cite[e.g.,][]{hullman2015content,kennedy2018feeling,peck2019data} we discuss four drivers for expressing personal experiences and perspectives and seek to understand what they add to the discourse around data visualization. 

\textbf{Expressing opinions}---We observed that reactions expressing opinions, i.e., points of view or judgements about a topic, are an inherent part of the sensemaking process, as they act as an intermediary between the presented information and a viewer's self-perception. This self-perception can be shaped by cultural values or political views,
(e.g., \reaction{[This] is a perfect encapsulation of Trump's opportunism, his hypocrisy and lack of self-awareness} \vistag{golf})
but also relate to personal preferences, e.g., how people react to the presented ratings of a TV show they enjoyed. Expressions of opinions vary in their expressivity including typical phrases like ``I think / love / hate [...]'' or ``in my opinion [...]''. Paired with a scope of topic or insight---specifically when talking about their own observations---they further indicate that people refocus on their own position and juxtaposing it to what the data suggests when making sense of it. Commenters are well aware of this dynamic that prefers personal relation over information: \motivation{I believe my comment was upvoted because others agreed with it, not because it offered any insight or analysis.}

\textbf{Matching knowledges}---Other reactions demonstrate how people compare extracted information against their own existing knowledge. Oftentimes, this simply results in people stating what they learned through observation and interpretation. In case of a mismatch between presented and existing knowledge, we observe two possible reactions. We found several reactions that imply how the new information challenges previous assumptions, often accompanied by an expression of astonishment (\reaction{Bernie Sanders is 78 years old with a NW of only 2.5 million? That's honestly pretty surprising to me [...]} \vistag{net-worth}. If however, the viewer is not convinced by the presented data, they express their own knowledge in the form of critiques, motivated by not letting something they perceive as wrong remain uncommented:
\motivation{I have a compulsion for calling out bullshit by using my experience and education}. 

\textbf{Sharing emotions}---Redditors post emotional reactions that are evoked by the visualization. While many reactions contain emotions, they do not typically carry the central message of the reaction, but often
set the underlying tone of a reaction. People prelude their reactions with how a visualization makes them \textit{``feel''}, or what their \reaction{gut feeling} \vistag{garbage} about the data is.
Some reactions however have feelings or emotions as a first class citizen:
\reaction{I'm simultaneously horrified yet excited to see how this looks [...]} \vistag{cases}. In the survey, one participant points out how expressing emotions is the main driver of her reaction: \motivation{I can't really explain my motivation other than wanting to share how this phenomenon 'feels', even if feelings 
[do not equal] science.} Another participant states: \motivation{This was about the hard feelings I still have [...] and less about data as a motivation to post.} This emphasizes the role of emotional statements as a counter-weight against the typical impression of data and visualization to be rational and objective~\cite{d2016feminist}.

\textbf{Claiming interest}---We found some expressions that were shaped by a viewer's personal interest, which is clearly stated in their motivation: \motivation{I appreciate data visualization as someone with a background in the sciences, and also have an interest in politics, and so I felt motivated to comment [...]}. When formulating proposals, people often asked for an adaptation that they can better relate to personally, for example, in order to \motivation{satisfy a curious itch}. This might be a change of the visualization in order to better understand how the presented information affects them personally (\reaction{Can we get another chart that outlines the reduction of state and federal funds dedicated to higher education? I know my state only contributes to less than 10\% of the state's higher education system} \vistag{tuition}),
or the adaptation of new data, for example, to display the ratings of a TV Show they watched and want to learn more about. 
In the survey, a participant described his motivation to post a clarification with a \motivation{genuine interest to get an apolitical answer [...]} in order to learn more about particular aspects that were not further described in the visualization. People also ask for clarifications to express their \motivation{Desire for more than one perspective on a dataset}, which can be met by adding anecdotal evidence.

\textbf{Adding anecdotal evidence}---When considering a viewer's personal involvement in the visualizations, testimonies get to the heart of what the previous drivers already indicate. They are expressions of inherently personal, sometimes even intimate relations that complement the visualization with first-hand experiences. Against the background that visualizations can depict data that may be specific to a single country, time-period or demographic group, these reactions offer subjective perspectives that can help the audience to better relate to the data. Several participants confirm this motivation: 
\motivation{I was looking to provide a human story behind the data. The bushfires in Australia earlier this year were devastating and my comment was intended to describe how the fires being discussed in the data is beautiful post was affecting millions of people} 
 or 
\motivation{I am a regular reader of Reddit, and golf a lot. The data was interesting to me and I wanted to add some context to it for those that may not golf and do not quite understand if the rounds of golf represented are a lot or not very many. If an activity is foreign to someone a number matters little to them. Context or comparison to better understand the number can be helpful.}
\subsection{Informing the Public Discourse}

We found that many personal reactions, though motivated by internal reasons, were meant to contribute to the public discourse and collaborative sensemaking among community members. 
For example, while some reactions only speak to the visualization author (\motivation{To compliment the OP and give constructive feedback}, \motivation{encouragement [...] and asking for more information}). Other comments aimed to add to a public discourse more broadly: \motivation{posting my opinion to a wide audience}. Below, we characterize four motivations of people to express themselves in public. Considering that reactions are also motivated by personal expression it is important to note that these drivers are not mutually exclusive with the personal ones reported in section 5.1. 

\textbf{Participate in public sensemaking}---
Several participants explained their motivation to share their existing knowledge or gained insights with others as constructive and objective conclusion to share facts, findings, and implications from the visualization with others (\motivation{[...] understand and share the implications of the data}; \motivation{share an interesting conclusion [...]}; \motivation{share an observation/trend}).

\textbf{Adding one's own story}---Personal experiences can constitute a crucial quality of the reactions to data visualizations we studied. The prospect of providing complementary human perspectives onto the data is particularly relevant for the discourse to humanize data, complement findings, \motivation{providing local or 'inside' information}, or to \motivation{contextualize the data for others[...]. I wanted people remember or know what it was like.}.

\textbf{Rectifying what's wrong}---Correcting wrong information goes beyond the previously described sensemaking. Similar to previous work~\cite{hullman2015content}, we observed how people express critique about the topic, data, or visual representation: \motivation{Data is
often used to justify one side of things. Unfortunately data can be manipulated to make concepts, ideas, etc. seem better or worse [...]}.
While correcting old or introducing new information is an important motivation, we only see few cases that use and link to verifiable sources. In some cases, the source is named but not linked---or different authors provide different ``correct'' numbers fixing the data. The majority of reactions 
claiming that some of the presented data is wrong and suggest other data
remain unverified and are intertwined with personal perspectives.
Eventually, critique on the visual design is specifically important and points to increasing awareness of slants and biases in visualization: \motivation{A lot of information is presented in a biased way, where the numbers are accurate but are portrayed in biased light. I try to rectify that}.




\textbf{Sparking discussion}---Reactions can be an invitation for others to participate. Participants express reactions in order to further \motivation{engage with the community} and trigger more follow-up reactions: they want \motivation{to bring attention to a particularly interesting data event and provide a starting point for further discussion} and feel that the statements they make \motivation{furthered the discussion around the topic}. People are curious what other's make of their observations and opinions: \motivation{I thought it'd be interesting to see how people would react to it [and] if others felt the same [...]}. 
\\


%

Besides personal and public motivations, a third group of motivation can be summarized \motivation{acquir[ing] upvotes}, or having fun by contributing jokes: \reaction{sparkles! I love sparkles} \vistag{cases}.



\section{Discussion}
\label{sec:discussion}

In this section, we summarize and discuss our findings, 
reflect on our methodology, discuss implications
for interface design to support engagement with visualizations, and relate our findings to other findings in the literature.

\subsection{Main Findings}

We found a great \textbf{diversity of reactions} in the visualization-focused online community \texttt{/r/dataisbeautiful}. Previous classifications~\cite{vanhulst2018descriptive,vanhulst2018designing} and interfaces~\cite{heer2007voyagers,willett2011commentspace} primarily focused on annotations---a different, yet arguably comparable kind of audience-authored reaction---as tools for collaborative sensemaking. In contrast, reactions on Reddit serve a wider variety of purposes and express not only analytical insight, but also personal perspectives, critical questions, and additional information. 
Furthermore, the comments themselves constitute an integral part of the visualization community on Reddit and as such they also exhibit humor, sarcasm, and banter.
The dynamic between the visualization and the people that comment on it is oftentimes harmonic and sociable, for example, when expressing observations or evoking memories, but can also be tense and antagonistic, especially when critical comments and opinions are expressed. 

We were interested in understanding how an open and potentially heterogenous community reacts to visualizations, differing from from lab studies and specific communities (such as university students and employers of a tech company) and a wide range of different types visualizations. Our results confirm findings by Heer et al.~\cite{heer2007voyagers} and reveal additional categories. Our taxonomy is meant to capture ‘exclusive’ categories in order to provide a better picture of individual types of reactions. Finally, we are most interested in how people report on their personal relation and their motivations in contributing to commenting, something entirely missing in the previous studies.


For example, we found various \textbf{levels of sensemaking} across almost all reaction types.
While observations can be associated with low engagement in Mahyar et al.'s taxonomy~\cite{mahyar2015towards}, reactions such as hypotheses, clarifications, and conclusions indicate a higher level of engagement,  often based on personal experience, wider contextual knowledge, or profound interest. A specific type of comments added comparisons and other information to help translating the insights of a visualization from one context to another. Eventually, critiques can be seen as a distinct form of sensemaking, given that they imply some sort of personal cognitive effort that includes understanding an issue, comparing it with previous observations or guidelines, finding a point of disagreement, and expressing this disagreement intelligibly. In many cases, the critiques we found were highly constructive, suggesting the means to overcome a particular weakness. 

The \textbf{role of personal perspective and experience} is prevalent across all types of reactions and ranges from expressing opinions and emotions to contributing personal experiences (testimonies) and adding or providing evidence.
We also found that critique is not necessarily expressed from a neutral or analytical perspective: more than one third of critical reactions have a personal notion to them. This supports many of the findings by Peck et al.~\cite{peck2019data} and also shows that personality does not only shape perception, but also the sensemaking and communication after viewing a visualization. Confirming Kennedy et al.'s~\cite{kennedy2018feeling} observations about the role of emotions in engagements with data, we found multiple instances that expressed the relevance of people's \textit{feeling of numbers}. Eventually, we identified several \textbf{drivers for engaging with visualization}, including personal motivations and contributions to the public discourse.

\subsection{Reflecting on Methodology and Data}

Our findings are characteristic for Reddit, specifically the \texttt{/r/data\-isbeautiful} community. Reddit is featuring non-interactive content, other than e.g., many contemporary news outlets and their visual stories. Besides the ensuing discussions, Reddit content comes without additional information, e.g., such as in Hullman et al.'s study~\cite{hullman2015content}. While related studies on personal perception~\cite{peck2019data} and feelings~\cite{kennedy2018feeling} have been carried out in co-located, synchronous workshops, or interviews, our study focused on an online setting. For the most part, we were able to support existing results and deepen our knowledge of the audience behavior when reacting to visualizations and their motivation behind it. 

We deliberately \textbf{ignored discussion threads}, i.e., the nested replies to comments, in order to focus on the primary reactions to the visualizations. However, we often resorted to replies in order to understand and agree on the meaning of single comment as these replies often clarified the initial intention of a comment. Future work should consider these discussions to better understand the audience's perception of reactions, for example, to find out how successful testimonies are in humanizing the data or determine whether people change their mind based on critiques. The reaction types provide a structured basis for such an endeavour and can be used to further analyze how discussions evolve.

To account for the complex nature of the expressions we encountered, 
we \textbf{iteratively adjusted the coding scheme}. Comments contained traces of different reaction types and this ambiguity required intense discussion among the authors.
Correctly and consistently interpreting a reaction depends on inside knowledge about the topic, the specific online community, and the used jargon and abbreviations. Some of this is due to the varying familiarity with Reddit and the \texttt{/r/dataisbeautiful} community among the authors. 
Still, some comments could not be classified due to potential irony, missing context, or where the reaction would not fit in any of the ten categories.

More research is necessary to \textbf{generalize our findings} and determine whether they are specific to one online community or whether they apply to other platforms, communities, and even co-located (uncontrolled) settings. Our framework can serve as a reference point to understand how reaction types, their quality, and quantity change with context (e.g., online, situated, synchronous, asynchronous~\cite{isenberg2011collaborative}), audience, media (e.g., pen and paper, data physicalization, interactive
), and goals of engagement (e.g., open, entertainment, collaborative sensemaking, design crits). Evidently, more research is necessary to understand these specific parameters that influence human engagement with data visualizations. 

Our reactions are \textbf{based on textual comments}. Other sources of textual comments could include forums~\cite{diehl2018visguides}. Other means of data capturing, for example through audio recordings, interaction, pointing, drawing, or specific functions in existing annotations interfaces, can provide complementary information while at the same time requiring specific analysis methods.

\subsection{Design Implications and Future Work}
Beyond the often discussed cases for engagement exploration and communication, our work informs mechanisms of questioning and contestation (\textsc{critiques}, \textsc{clarifications}), user contribution (\textsc{tes\-timonies, additional information, proposals, opinions}), and more sensemaking tasks (\textsc{observations}, \textsc{hypothesis}, \textsc{conclusion}). 
Informed by these reactions (\ref{sec:reactions}) and drivers (\ref{sec:personalpublic}) for personal contributions,this section discusses design implications for user interfaces, complementing previous implications by Heer et al.~\cite{heer2007voyagers}.

\textbf{G1}---Allow for \textbf{type-specific annotations} to capture the richness of observers reactions and to support participation and discourse around data. Such rich annotations can entail audience-driven labeling of elements and regions of a visualization, highlighting relations between elements, linking reactions to data points, or drawing on the visualization. While some techniques are present in existing systems (e.g. Lyra, CommentSpace, Sense.us), our typology can be used to better tailor annotation features to suit the specific needs of individual reaction types. For example, interfaces could provide features to collect and categorize \textsc{opinions} such as done using Twitter for general discussions~\cite{huron2013polemictweet} and link them to the data. We can imagine visual polling interfaces in which user reactions drive visual properties of elements that are put to the vote
, e.g., the roughness of a visual shape or a highly controversial opinion. Interfaces for data contestation could allow people to \textit{rectify whats wrong} and provide features to express a \textsc{critique} of data points or the quality or provenance of the data set, and add details on the annotator’s confidence or the provided \textsc{additional information} as evidence. \textsc{clarifications} and \textsc{proposals} aim to improve or re-create visualizations as proposed by Hullman at al.,~\cite{hullman2015content}, raising questions about data access and visualization creation tools. 

\textbf{G2}---Supporting \textbf{personalized stories and visualizations} can capture an audience's \textsc{testimonies} and display them within the context of a given visualization.
Driven by people's willingness to \textit{provide anecdotal evidence} and \textit{add their own story}, this allows for a juxtaposition of the data's bird's-eye perspective on a topic and a local, situated reaction to it and thus provides a more convincing narrative as proposed by one study participant:
\motivation{visualization is powerful - it's even more powerful when supplemented by real experiences}.
We imagine interfaces to provide personal testimonies in a more participatory and empowering way than `just` annotating and commenting on a given visualization. Challenges are posed by allowing for easy authoring of such stories as well as storage and navigating these stories. Eventually, such stories can include not only text, but other media such as photos, oral histories~\cite{maharawal2018antieviction}, customized visualizations, or data comics~\cite{zhao2015data,bach2017emerging}. This can help to include external information and adapt a given visualization to different contexts (e.g., using alternative geographic maps). 

\textbf{G3}---\textbf{Visualizing the discourse} around a given visualization can help analysts and an audience understand controversial parts and aspects of a visualization and the data displayed. For example, similar to hotspots in a document~\cite{hill1992edit}, a visualization could display a heatmap to highlight parts with many reactions, broken down to different reaction types, e.g. \textsc{opinions}, \textsc{contestations}, \textsc{clarifications}, \textsc{critiques}, \textsc{observations} or, as suggested by Heer et al.~\cite{heer2007voyagers}, keyword tagging.
Complementary analytical charts could show specific insights and trends about reactions such as reaction type, frequency over time, types of \textsc{opinions}, or terms mentioned in \textsc{critique} posts using techniques from NLP and sentiment analysis. Showing which and how elements of a data set or visualization are reacted to could inform both further audience engagement as well as a detailed analysis into the content of the discourse. 
An eminent challenge here is again to design easy-to-use interfaces to submit reactions and to navigate an existing discourse and its reactions.

\textbf{G4}---Open discourse and annotation requires \textbf{guidance and moderation}. \textbf{Guidance} should help participants express reactions and participate a discourse. For example, besides explanatory messages, visualizations of discourse (G3) can provide cues and affordances to an ongoing discourse~\cite{hill1992edit} and provide for \textit{social navigation cues}~\cite{heer2007voyagers}. Scoring systems could allow audiences interacting with the discussion itself~\cite{kauer2019participatory}. \textbf{Moderation} must prevent vandalism and misinformation as well as prevent discussions from going off-topic. Also, moderation can help finding agreement and resolving conflicts for \textsc{contestations} and \textsc{critiques}, eventually updating a given visualization or data alongside respective notifications to inform the audience of such updates.

\textbf{G5}---\textbf{Gratify people’s participation}. While one goal of any (visualization-based) discourse should be to enrich an audiences' understanding of the data (see Sec 5, \textit{matching knowledge, add anecdotal knowledge, participate in public sensemaking, sparking discussion, rectifying what's wrong}), we found many other motivations for people to engage with visualizations (express opinion, claim interest, share emotions, express humor) which could be leveraged by social media platforms in to attract attention to a topic and give the audience an enriched channel of expression and participation. More simple motivations can include attention to one's comments or posting humoristic comments. One challenge here is how to get people engaged in intellectually demanding reactions such as \textsc{critiques} and \textsc{contestations}, or suggestions on visualization redesign.

\section{Conclusion}


We see a huge research potential in providing more expressive and inclusive interfaces for engaging with and discussing content in data visualizations; this includes both online as well as offline media. 
To inform these interfaces, this paper has investigated reactions to data visualizations on the social platform Reddit, seeking to better understand different types of reactions and the drivers that motivate people to express them in a public forum. We coded comments to popular posts containing data visualizations and developed a coding scheme of ten types of reactions. We extended prior classification systems and added personal relations to data and a wide variety of critiques on the presented visualizations and their data sets. In a subsequent online survey, we identified nine drivers of people to engage with a visualization and express themselves. We differentiated between drivers that share a rather personal motivation from those that primarily aim to inform the public discourse. Our results show a variety of ways for people to personally, analytically, or critically connect with data when they are visualised in online media. Our work demonstrates people's willingness to engage and put effort into presenting personal perspectives to a public forum. From our findings, we deduced five design implications to inform future interfaces for participatory data visualization systems that support a range of different reaction types.

\section*{Acknowledgements}
\textit{We thank the reviewers for their valuable feedback and especially the program subcommittee member who helped shepherding our paper. We also thank the survey participants and the /r/dataisbeautiful community for their efforts to create, discuss, and moderate visualization content.}

\bibliographystyle{ACM-Reference-Format}
\bibliography{sample-authordraft}
\end{document}